# The Large-Scale Polarization Explorer (LSPE)

The LSPE collaboration: S. Aiola[a], G. Amico[a], P. Battaglia[b], E. Battistelli[a], A. Baù[c], P. de Bernardis[a], M. Bersanelli[b], A. Boscaleri[d], F. Cavaliere[b], A. Coppolecchia[a], A. Cruciani[a], F. Cuttaia[e], A. D' Addabbo[a], G. D'Alessandro[a], S. De Gregori[a], F. Del Torto[c], M. De Petris[a], L. Fiorineschi[f], C. Franceschet[b], E. Franceschi[f], M. Gervasi[c], D. Goldie[g], A. Gregorio[h,p], V. Haynes[i], N. Krachmalnicoff[c], L. Lamagna[a], B. Maffei[i], D. Maino[c], S. Masi[a], A. Mennella[c], Ng Ming Wah[i], G. Morgante[e], F. Nati[a], L. Pagano[a], A. Passerini[c], O. Peverini[l], F. Piacentini[a], L. Piccirillo[i], G. Pisano[i], S. Ricciardi[e], P. Rissone[f], G. Romeo[m], M. Salatino[a], M. Sandri[e], A. Schillaci[a], L. Stringhetti[n], A. Tartari[c], R. Tascone[l], L. Terenzi[e], M. Tomasi[n], E. Tommasi[o], F. Villa[e], G. Virone[l], S. Withington[g], A. Zacchei[p], M. Zannoni[c]

[a]Dipartimento di Fisica, Sapienza Università di Roma, P.le A. Moro 2, 00185 Roma, Italy
[b]Dipartimento di Fisica, Università di Milano, Via Caloria 16. 20133 Milano, Italy
[c]Dipartimento di Fisica, Università di Milano Bicocca, Piazza della Scienza 3, 20126 Milano, Italy
[d]IFAC-CNR Via Madonna del Piano, 10 50019 Sesto Fiorentino (FI), Italy
[e]IASF-INAF Via Gobetti 101, 40129 Bologna, Italy
[f]Dip. Meccanica e Tecnologie Industriali, Univ. di Firenze Via S. Marta, 3, 50139 Firenze, Italy
[g]Cavendish Laboratory, University of Cambridge, JJ Thomson Avenue, Cambridge CB3 0HE, UK
[h]Physics Department, University of Trieste via A. Valerio 2, 34127 Trieste, Italy
[i]Jodrell Bank Centre for Astrophysics, University of Manchester, Macclesfield, SK11 9DL, UK
[l]IEIIT-CNR, Corso Duca degli Abruzzi 24, 10129, Torino, Italy
[m]Istituto Nazionale di Geofisica e Vulcanologia, Via di Vigna Murata, 605, 00143 Roma, Italy
[n]IASF-INAF, Via Bassini 15, 20133 Milano, Italy
[o]Agenzia Spaziale Italiana, Viale Liegi 26, 00198 Roma, Italy
[p]OAT-INAF, Via G.B. Tiepolo 11, 34143 Trieste, Italy

## Abstract

The LSPE is a balloon-borne mission aimed at measuring the polarization of the Cosmic Microwave Background (CMB) at large angular scales, and in particular to constrain the curl component of CMB polarization (B-modes) produced by tensor perturbations generated during cosmic inflation, in the very early universe. Its primary target is to improve the limit on the ratio of tensor to scalar perturbations amplitudes down to $r = 0.03$, at 99.7% confidence. A second target is to produce wide maps of foreground polarization generated in our Galaxy by synchrotron emission and interstellar dust emission. These will be important to map Galactic magnetic fields and to study the properties of ionized gas and of diffuse interstellar dust in our Galaxy. The mission is optimized for large angular scales, with coarse angular resolution (around 1.5 degrees FWHM), and wide sky coverage (25% of the sky). The payload will fly in a circumpolar long duration balloon mission during the polar night. Using the Earth as a giant solar shield, the instrument will spin in azimuth, observing a large fraction of the northern sky. The payload will host two instruments. An array of coherent polarimeters using cryogenic HEMT amplifiers will survey the sky at 43 and 90 GHz. An array of bolometric polarimeters, using large throughput multi-mode bolometers and rotating Half Wave Plates (HWP), will survey the same sky region in three bands at 95, 145 and 245 GHz. The wide frequency coverage will allow optimal control of the polarized foregrounds, with comparable angular resolution at all frequencies.

**Keywords:** Cosmic Microwave Backgorund, Polarization, Polarimeter, Bolometer, Radiometer, Stratospheric Balloons

# 1. INTRODUCTION

CMB polarization has been measured by several experiments (see e.g.[1-6]) over a wide range of angular scales. The results are consistent with the polarized signal produced through Thomson scattering by the same scalar perturbations producing CMB temperature anisotropy (E-modes). Improved measurements are expected by forthcoming experiments (e.g.[7-10]): the target is to measure the rotational component of CMB polarization (B-modes) which, at large angular scales, represents a unique signature of the tensor perturbations produced during inflation[11]. An experiment measuring B-modes must detect the variance of these fluctuations, which results from the contributions of all the detected multipoles. For full sky coverage

$$\langle \Delta T_{BB}^2 \rangle = \frac{1}{4\pi} \sum_{\ell=2}^{\infty} (2\ell+1) C_{\ell,BB} B_\ell^2 \qquad (1)$$

where $C_\ell$ is the angular power spectrum of the fluctuations, $B_\ell^2$ is the beam window function, i.e. the response of the telescope to different multipoles. Most of this signal comes from large angular scales (low multipoles, see Fig.1).

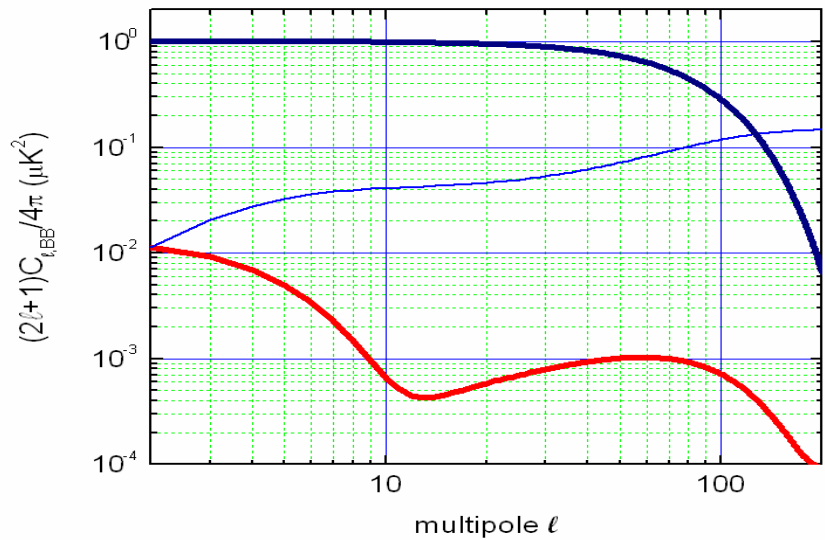

Figure 1. The bottom thick line represents the contribution from each multipole to the total mean square fluctuation of the tensor component of CMB polarization (B-modes, assuming a tensor to scalar ratio $r = 1$). The bump at small multipoles is due to photons last scattered during reionization, while the second bump is due to photons from z=1100. The thin line is the cumulative of the B-modes, i.e. the variance measured by an experiment sensitive from multipole 2 to a given multipole $\ell$ (see eq. 1). The top thick line represents the beam function $B_\ell^2$ for an experiment with a 1.5° FWHM Gaussian beam: despite of the coarse angular resolution such an experiment collects most of the polarization signal from B-modes.

The importance of this tiny signal has been stressed widely in the literature. A detection of B-modes in the CMB polarization field would represent the final confirmation that inflation really happened[11,12]. The detected tensor to scalar ratio $r$ would constrain the energy-scale of inflation, according to the relation[11-15]

$$E = 3.3 \times 10^{16} r^{1/4} \, GeV \qquad (2)$$

thus allowing for the first time an observational incursion into the physics of extremely high energies. Indeed, no conceivable experiment in a terrestrial laboratory can probe such ultra-high energy regime[13]. Moreover, high precision CMB polarization measurements are essential to study CMB lensing and thus constrain neutrino masses[16-18].

There is no clear theoretical forecast for the value of $r$. For this reason it is very important to carry-out pathfinder experiments, like LSPE, before that an extremely expensive third-generation space mission devoted to the final CMB polarization measurement is decided.

It is, however, very difficult to measure sky polarization at large angular scales, because this requires wide sky coverage, high stability of the instrument performance and of the atmosphere over large volumes, reduction and control of telescope sidelobes over a wide range of angles. Moreover, the expected signal is very small with respect to E-modes and local foregrounds.

In this paper we describe the Large Scale Polarization Explorer (LSPE), a long-duration balloon experiment exploiting the environment of the stratosphere during the polar night, to measure B-mode CMB polarization at large angular scales. To do this, the experiment spins continuously at 3 rpm, covering a large fraction of the sky in a single balloon flight. Two independent instruments are aboard of the LSPE: the STRatospheric Italian Polarimeter (STRIP)[19] and the Short Wavelength Instrument for the Polarization Explorer (SWIPE)[20]. The two instruments are accommodated on a common frame, which provides flight control, power supply, sky scanning, and communication services. The experiment has been designed to improve over previous efforts in three areas:

- Sensitivity: the sky survey sensitivity is improved using arrays of photon-noise limited detectors (multimode for SWIPE). This is discussed in detail in two companion papers[19,20].

- Systematic effects: their mitigation and control have been improved using several levels of modulation, and a polarization modulator (a HWP in the SWIPE instrument, while the STRIP coherent instrument uses correlation radiometers). This is also discussed in detail in[19,20]. Both instruments are sensitive to the W band, using orthogonal technologies and thus providing an important additional check for systematic effects.

- Foregrounds: the frequency range covered by the two instruments brackets the region where the ratio between polarized CMB signal and polarized foregrounds is maximum. This allows us to use efficiently component separation techniques.

Polarized foreground emissions are likely to be the main limitation for B-modes detection at large angular scales. Our current knowledge of diffuse polarized foregrounds mainly comes from WMAP data (covering the 23-94 GHz frequency range). New data from the Planck satellite will soon become available, allowing more precise estimates in this range, extending our knowledge above 100 GHz. The main diffuse polarized emissions are synchrotron emission and thermal dust emission from non-spherical grains. Diffuse synchrotron emission is polarized around 10-20% on the average[21, 22], the polarization of thermal dust is expected to be about 5% [23]. Free-free and anomalous dust emission should be essentially unpolarized (current upper limits to their polarization are around 3%)[24]. Even masking the most contaminated regions of the sky (P06 WMAP mask, excluding 30% of the sky), the contribution of polarized foregrounds in the cleanest WMAP frequency band (61GHz, V band) is at least one order of magnitude higher than the CMB E-modes at large angular scales ($\ell<20$). For an experiment like LSPE, having a native coverage around 25% of the sky, an aggressive sky masking is not a viable option, as it will increase the cosmic variance term which is already relevant. It is clear that an accurate foreground removal is needed.

In the last few years several data analysis techniques have been proposed (e.g.[25]), exploiting the multi-frequency information to reduce foreground contamination on the CMB emission (foreground cleaning methods) or to disentangle the single components (component separation). Most of these methods perform a linear mixture of the instrumental channels, with weights designed for different trade-offs between instrumental noise and foreground contamination. The best strategy clearly depends on the instrumental specifications and should be optimized towards the scientific goals of the experiment (e.g.[26]). A detailed simulation effort specifically tailored to LSPE is underway.

Preliminary results are illustrated in Fig.2. The figure reports the expected errors on angular power spectra in two different cases. In black, the maps have been combined using the minimum variance approach, thus weighting only by the inverse noise. In gray the maps are combined using the minimum foreground approach, thus the weights take into account foreground minimization. In doing that, an imperfect knowledge of the foregrounds frequency scaling is assumed. In particular, with an optimistic approach, we assume to know with good accuracy the spectral index of the synchrotron component (3.00+0.01) and the spectral index for the dust (1.700+0.006). For both cases, we plot three information: the residual noise (square dots), the residual foreground (triangular dots), and the sum of the two (continuous line). Cosmic variance is ignored, as in the case of no B-modes. From the figure it is evident how the foreground minimization improves the result with respect to the noise minimization, in particular at large angular scales.

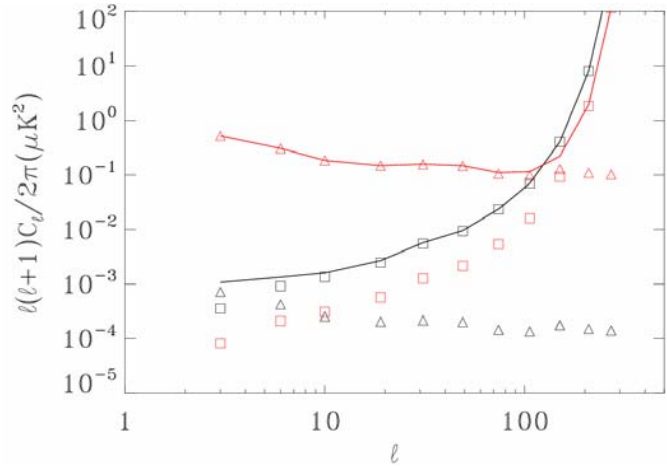

Fig.2: Forecast for B-modes detection by LSPE . Bottom solid line: total error for a component separation weighting scheme that minimizes foregrounds residuals. Top solid line: total error for the minimum variance weighting scheme. The total errors are the sum of the noise (triangles) and foreground (squares) errors. A weighting scheme that minimizes foregrounds residuals will lead to some amplification of the noise (as evident in this plot) but remains a reasonable choice, since it reduces the impact of foregrounds by orders of magnitude. See text for details.

## 2. BALLOON FLIGHTS IN THE POLAR NIGHT

Long duration (2 weeks or more) stratospheric balloon flights are routinely performed every year by NASA-CSBF in Antarctica, during the summer season ( see e.g.[27]). These flights offer continuous operation of heavy payloads (up to 2 tons) in the stratosphere (at altitudes around 35 – 40 km) for more than 2 weeks (and up to 1 month). The successful ATIC[28], BLAST[29] and BOOMERanG[30] are good examples of experiments exploiting this opportunity. In the Arctic, an exploratory program has been carried out in collaboration with the Italian Space Agency, demonstrating the feasibility of long duration flights launched from Longyearbyen (Svalbard Islands)[31] (see Fig.3).

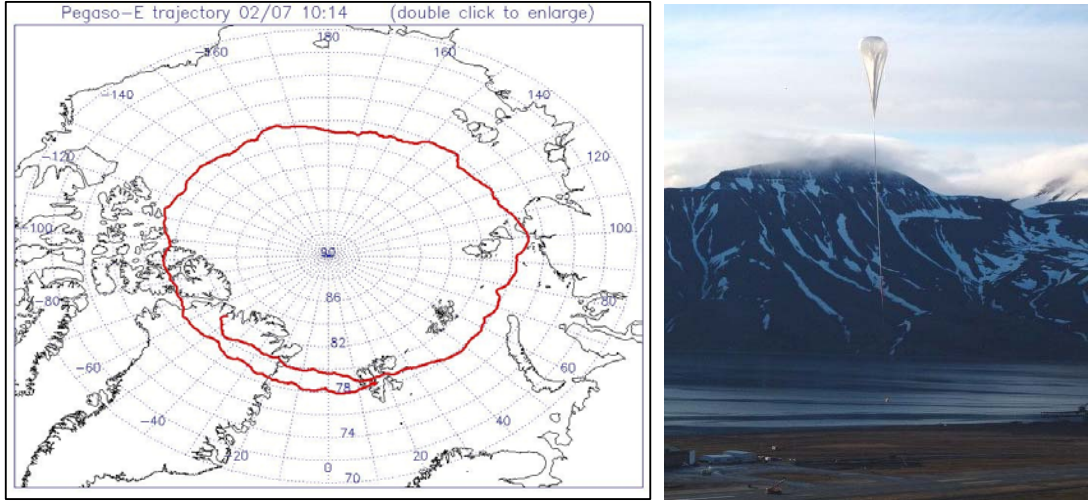

Figure 3. **Left:** Ground path of one of the test flights performed by ASI to test the stratospheric circulation near the North Pole. This flight was performed in summer. **Right:** Launch of a heavy-lift balloon from the Longyearbyen airport (Svalbard Islands, latitude 78°N).

Testing of night-flights has also been performed from the Nobile-Amundsen base in Ny-Alesund (Svalbard), demonstrating the eastward trajectory of the stratospheric probes during the polar night[32] (see Fig.4).

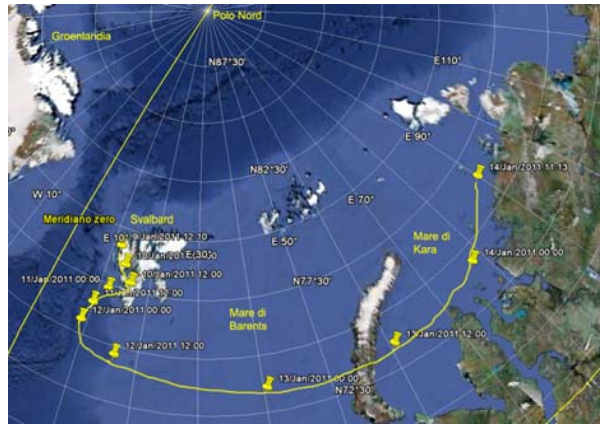

Figure 4: Ground path of a small pathfinder test flight performed in January 2011, in the middle of the polar night. The eastward trajectory is evident.

At lower latitudes, polar night flights have been performed by CNES from the Kiruna base of the Swedish Space Corporation, with recovery of the payload in Russia, like in the case of the Archeops experiment[33]. In our case we plan to launch the LSPE from Longyearbyen in the polar night, for a long duration (15 days nominal) flight around new-year. Recovery will be in Greenland. In case of successful recovery, we will consider a second flight in the southern hemisphere.

There are specific technical problems to face for such a mission, in addition to those of a normal day-time flight. The first one is the payload thermal management. The temperature of the stratosphere during the polar night is around -80$^o$C, and there is no solar radiation available to warm-up the payload. So all the subsystems must be able to withstand these harsh conditions. The most important impacts are on the electronic systems, which must be thermally insulated from the environment to achieve self-heating conditions, on vacuum seals, which must be manufactured in indium or in special elastometers, on mechanical actuators, whose play and lubrication must be specially designed for low temperature and low pressure. The second technical problem is the supply power for the experiment and the telemetry. With a typical power requirement exceeding 700 W for 15 days, an electrical energy storage close to 1 GJ is needed. The simplest solution is to use lithium batteries, which feature high energy density and can operate in the vacuum and at very low temperatures. Fuel cells represent another possibility, but still not validated for space applications. In both cases the cost is significant. Due to their very small internal resistance, lithium batteries should be placed inside the same insulated box containing the powered electronics, to reach a temperature above 0$^o$C and maintain most of their nominal capacity.

Data transmission is also problematic. LSPE will produce a raw data rate of about 400kbps, which is entirely stored on-board on solid state disks. Essential housekeeping information is transmitted though the iridium network, to check the performance and the evolution of the observing program. The scientific data set is too large for this communication system. We plan to have line-of-sight telemetry at full rate during the first day of the flight, and then to have data-dumps when the system flies over selected locations hosting dedicated receiving stations.

## 3. THE LSPE INSTRUMENTS

STRIP is an array of coherent polarimeters, whose main target is the accurate measurement of the low-frequency polarized emission, dominated by Galactic synchrotron. Its design is described in detail in[19]. We summarize in Table 1 below its performance, comparing it to the Planck-LFI[34-36]. The Q band array improves the Planck sensitivity by a factor of 2, while the small W band array provides a common channel with SWIPE for internal crosscheck of systematic effects.

|  | **PLANCK LFI** | | | **STRIP** | |
|---|---|---|---|---|---|
| Frequency (GHz) | 30 | 44 | 70 | 43 | 90 |
| Resolution (deg) | 0.55 | 0.47 | 0.22 | 1.0 | 0.5 |
| Sky coverage (%) | 100 | 100 | 100 | 18 | 18 |
| Obs Time (months) | 30 | 30 | 30 | 0.467 | 0.467 |
| Bandwidth (GHz) | 4.5 | 4.1 | 12.0 | 7.7 | 16.2 |
| N_horn | 2 | 3 | 6 | 49 | 7 |
| Tnoise (K) | 11.3 | 17.0 | 33.2 | 20 | 41 |
| Telesc. / window (K) | < 1 | | | 7 | 8 |
| Tsky (antenna) (K) | 2.068 | 1.804 | 1.382 | 1.822 | 1.113 |
| 1-s sensitivity ($\mu K \times s^{1/2}$) | 147 | 173 | 153 | 33 | 104 |
| **Delta Q(U) per 1.5° pixel (micro-K)** | 3.27 | 3.95 | 3.77 | 1.78 | 6.68 |
| **Improvement factor wrt Planck-LFI** | | | | 2.2 | 0.6 |

Table 1: Main characteristics of the STRIP instrument, compared to the LFI aboard of Planck.

SWIPE is an array of bolometric polarimeters. Its target is the accurate measurement of polarization of the galactic dust foreground and of the CMB. Its design is described in detail in[20]. We summarize in Table 2 below its performance, comparing it to the Planck-HFI[37]. The 95 band is the main channel for CMB polarization measurements, while the 145 GHz and 245 GHz bands will measure precisely the polarized emission from interstellar dust.

|  | **PLANCK – HFI** | | | | | | **SWIPE** | | |
|---|---|---|---|---|---|---|---|---|---|
| Frequency (GHz) | 100 | 143 | 217 | 353 | 545 | 857 | 95 | 145 | 245 |
| FWHM Resolution (arcmin) | 9 | 7 | 6 | 5 | 5 | 5 | 110 | 89 | 74 |
| Sky coverage (%) | 100 | 100 | 100 | 100 | 100 | 100 | 20 | 20 | 20 |
| Obs Time (months) | 30 | 30 | 30 | 30 | 30 | 30 | 0.467 | 0.467 | 0.467 |
| Bandwidth (%) | 33 | 33 | 33 | 33 | 33 | 33 | 25 | 25 | 25 |
| N_det (polarized) | 8 | 8 | 8 | 8 | 0 | 0 | 80 | 86 | 110 |
| Channel NET (uK s^1/2) | 25 | 31 | 45 | 140 | // | // | 1.9 | 1.8 | 1.9 |
| Integration/beam (s) | 33 | 20 | 15 | 10 | - | - | 660 | 415 | 225 |
| **Delta Q(U) (uK) on SWIPE beams** | 0.27 | 0.42 | 0.84 | 2.6 | - | - | **0.10** | **0.13** | **0.17** |
| **Improvement factor wrt Planck-HFI** | | | | | | | **2.7** | **3.2** | **4.9** |

Table 2: Main characteristics of the SWIPE instrument, compared to the HFI aboard of Planck.

## 4. THE LSPE CARRIER

### 4.1 THE GONDOLA

The two instruments are mounted on a common azimuth spinning frame, on tiltable platforms to allow elevation control. The line of sight elevation can be changed independently for the two instruments, in the range between 15° and 50°. The nominal survey will be carried out at elevations between 40° and 50°. Lower elevations will be used only to scan planets, the Moon, and other calibration sources.

To host the experimental equipment, a structure characterized by a low weight and a high stiffness is required. A first result of the design activity was a type of gondola with high level of reusability and exploitable for future multi-user experiments, presented in[38], where also the systematic approach used for the design process was briefly described. However, the large final mass estimate of the payload drove a second design phase, with maximum lightness considered as the most important objective. A preliminary sketch of this new light-weight structure is shown in Fig.5.

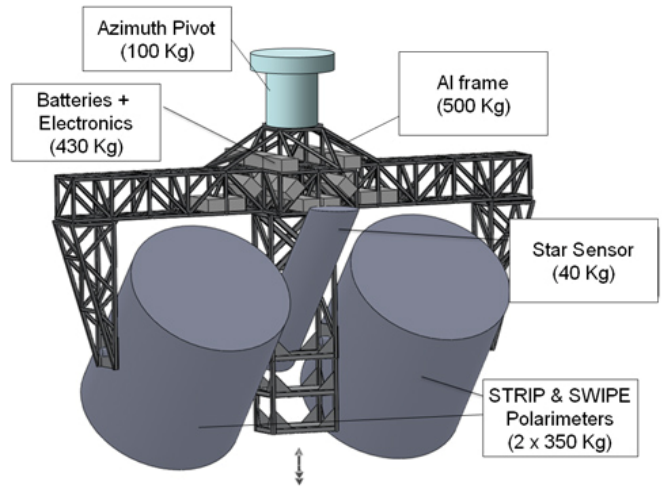

Figure 5. Preliminary sketch of the LSPE experiment. The width of the gondola for this configuration is about 5m.

The stiffness and strength specifications together with the dimensions of each structural element were verified using the Finite Element Method (FEM) from the ANSYS v.13 code. The simulations were performed first with an applied vertical mass acceleration of 10g, then with an acceleration equal to 5g and with an angle of 45° with respect to the pivot axis. These are the maximum accelerations loading the structure, during parachute opening, after termination of the flight. At the moment steel and aluminum alloys have been considered as construction materials. For further developments also composite materials will be investigated in order to reach more lightness and to reduce the rotational inertia of the structure.

At the current level of development, the total weigth of the payload is around 2.2 tons, which requires a 800000 m$^3$ balloon to fly at more than 37 km of altitude.

**4.2 THE ATTITUDE CONTROL SYSTEM**

The pointing directions of the two telescopes are controlled by the Attitude Control System (ACS). This is composed of a set of actuators, a processor and a set of attitude sensors. Its purpose is to produce the required scan of the sky and of the calibration sources, and to acquire attitude data precise enough to allow sub-arcmin reconstruction of the pointing during data analysis. This ACS is derived directly by systems we have developed for previous balloon missions[39-43].

- **Actuators:** The payload is decoupled from the flight-train by a Pivot system, including thrust ball bearings and a torque motor (Kollmorgen QT-6205) producing the azimuth rotation of the payload. The power required to rotate the payload at a constant rate of 3 rpm is due to the friction torque generated in the thrust bearings. With a payload weight of 2.2 tons, the friction torque is estimated to be around 0.2 Nm. The motor constant converting current into torque is 1.4 Nm/A, so we expect currents below 1A for the regime conditions and below 10A for the speed-up phase. Two independent elevation actuators, one for each instrument, control the relative angle between the inner frames and the outer frame. The total power needed to activate the motors is below 100W.

- **The Processor** is a PC104 unit which continuously acquires information from the sensors, evaluates the attitude and controls the current in the motors, thus controlling all the payload movements and sending pointing information to the data acquisition system and the telemetry.

- **Attitude sensors:** The attitude information are provided by several sensors, complementary one to each other. The absolute time and geographical position are given by a GPS unit. A three axis KVH E-Core 2000 fiber optic gyroscopes system can be rapidly sampled to get the angular velocity vector describing the azimuth motion and the pitch/roll oscillations with a 0.014°/s sensitivity. Absolute angular encoders will sample the elevation angles with 16 bit resolution. These sensors are used in real time by the processor to drive the actuators. A star sensor provides the information for off-line absolute pointing reconstruction as described below.

**4.3 THE FAST STAR SENSOR**

The star sensor is the main sensor for telescope pointing reconstruction, since it can provide the attitude information in an absolute reference frame. It produces a sky map where optical star signals can be identified if compared to a catalogue, thus giving the telescopes pointing with respect to the sky coordinates. It is based on the same solution we developed for the successful Archeops flights (which was carried out during the polar night, as planned for LSPE)[44,45]. This is based on simple and reliable components, providing high resolution and sensitivity with fast response, as needed in a fast sky scanning instrument. The telescope is composed of a single parabolic optical mirror, diameter 40 cm, f/4.5. A linear array of 46 photodiodes (Hamamatsu S4111) is placed in the focal plane together with its low noise readout electronics. The star sensor telescope points at constant elevation and spins together with the payload. The array orientation is perpendicular to the horizon, and each pixel is 4 mm wide in azimuth (or scan direction) and 1 mm wide in cross-scan direction. During the azimuth rotation, the star sensor covers a 1.46° thick stripe at the selected elevation angle. The sampling period will be around 5 ms. The system can detect about 50-100 stars per turn up to magnitude 6: more than enough to reconstruct the attitude of the payload with sub-arcmin resolution.

## 5. OBSERVATION STRATEGY

The LSPE payload will spin at constant speed (3 rpm) around the zenith axis, keeping the telescope boresight at constant elevation for long periods. This strategy is similar to the one adopted by the Archeops balloon experiment[45], and shares several commonalities also with the Planck satellite[7]. The elevation will be changed occasionally, in order to extend the full coverage, and to execute specific calibration observations, such as scans of planets or other calibration sources.

The two instruments have independent elevation movements. A night flight is a mandatory requirement for a spinning payload, to avoid contamination by sunlight. Winter polar flights are the only solution for this kind of strategy.

The spin of the payload, combined with the daily rotation of Earth, results in a sky coverage of up to 30% depending on elevation changes, at the expected latitude (78°N) of the flight. Table 3 illustrates the coverage of the two instruments for different elevation strategies.

The current baseline scanning strategy foresees a payload spin rate of 18°/s (3 rpm) and the coverage of the same sky area by the two instruments. Elevation changes will be operated once a day, at the same time for both instruments, to prevent mutual interference. Specific calibration observations will be executed in dedicated time slots by both instruments in parallel. Main targets of this phase will be Jupiter, the Moon, and the Crab nebula. Jupiter will be observed to map the main beam; the Crab nebula to calibrate the main axis of the polarimeters, since it is the most powerful polarized source observable at the LSPE frequencies and angular resolutions; the Moon can be used to map sidelobes, and the Moon limb can be measured to calibrate polarimeters. Jupiter and the Crab are observable by all detectors with limited elevation changes, while the Moon is observable only when it is opposite to the Sun. Table 4 illustrates sources culmination and sensitivity for a flight on January 1st 2015, at a latitude of 78° North. The sensitivity is calculated in terms of signal/noise, for a single sample, assuming a sample rate of 60 Hz. Except Jupiter, the other planets are too low to be observable, or too faint to provide a usable signal to noise ratio.

In Fig.6 we report the SWIPE 145 GHz Point Spread Function (PSF), overplotted on the expected data samples collected in co-scan and cross-scan coordinates. The simulation is based on the nominal scanning strategy, for one detector, with sample rate of 60 Hz, and represents only one day of observations. This figure demonstrates that the selected sampling rate results in a very dense sampling of the PSF, sufficient for an accurate reconstruction of all the details of the PSF.

Since Jupiter culmination is at 27° of elevation, using the nominal scanning strategy only a few detectors will scan over the planet. For the other detectors specific observations are required. We plan to dedicate one full day of the mission to the observation of calibration sources.

| Elevation | Coverage | Masked |
|---|---|---|
| SWIPE [30-40] | 31% | 23% |
| SWIPE [40-50] | 27% | 20% |
| SWIPE 35 | 24% | 19% |
| SWIPE 45 | 22% | 18% |
| SWIPE [30-50] | 35% | 26% |
| STRIP 45 | 27% | 20% |
| STRIP 30 | 33% | 24% |

Table 3: LSPE coverage for different sets of elevation changes. The first column reports the boresight elevation range in degrees for the two instruments. Second column, the full coverage. Third column, the coverage after masking the galaxy with the WMAP polarization mask.

| Source | Culmination (deg) | S/N per sample at 44 GHz | S/N per sample at 90 GHz | S/N per sample at 95 GHz | S/N per sample at 145 GHz | S/N per sample at 245 GHz |
|---|---|---|---|---|---|---|
| Moon | 30 | 37500 | 200000 | 2000000 | 700000 | 2000000 |
| Crab | 34 | 20 | 18 | 22 | 23 | 28 |
| Mars | 0 | 0.30 | 1.6 | 2 | 5.6 | 18 |
| Jupiter | 27 | 15 | 80 | 100 | 275 | 850 |
| Saturn | -6 | 1.4 | 7 | 9 | 24 | 70 |
| Uranus | 16 | 0.05 | 0.24 | 0.3 | 0.8 | 2.5 |

Table 4: Sources culmination angle, and sensitivity for a launch on Jan 1st, 2015. Sampling rate is set at 60 Hz. We assume full Moon, as it is when it is observable by LSPE. The Crab flux is based on the free-free spectrum reported in [46].

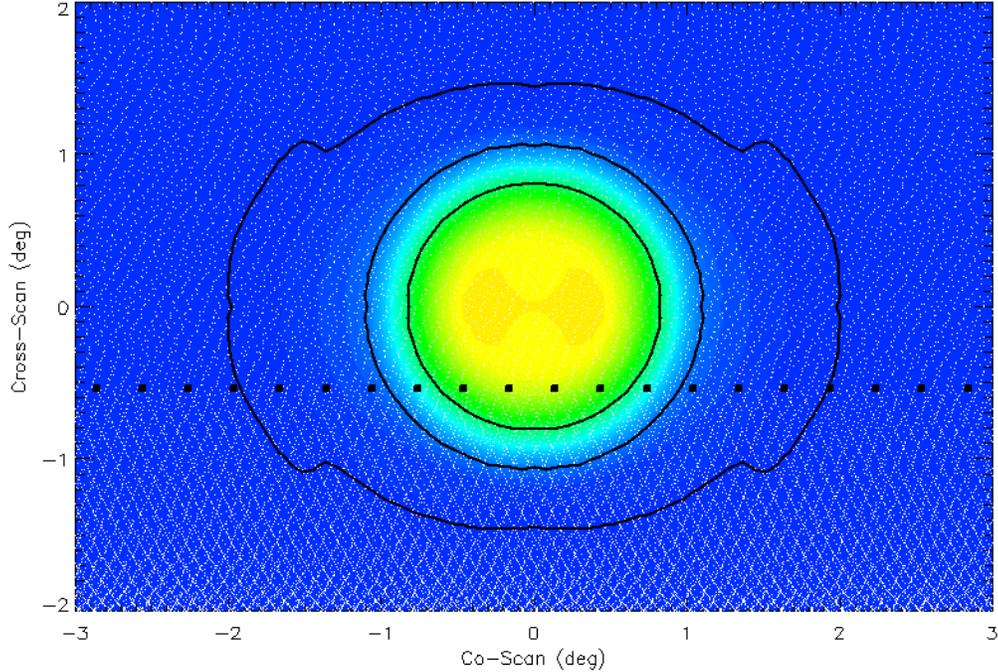

Figure 6: Simulation of data taken scanning over Jupiter. The colors represent the SWIPE 145 GHz point spread function (color levels) in co-scan and cross-scan coordinates. Contours are at 0.5, 0.1, 0.01 of the maximum.
The white dots represent the observed directions in one day of observations. The black squares represent the samples taken in a single pass over the source.

## 6. CONCLUSIONS

The LSPE balloon project is designed to meet the challenge of the next generation of CMB experiments, i.e., to search for the B-mode polarization component as a signature of an inflationary era in the very early universe. The LSPE instruments, the carrier and the observing strategy are optimized to probe large angular scales, where the B-mode signal is expected to be stronger. The extreme sensitivity of LSPE and the low amplitude of B-modes require a very clean separation of synchrotron and dust polarized components from the cosmological signal. LSPE is equipped with two instruments employing different technologies to ensure the broad spectral coverage needed to disentangle foreground contamination. A further challenge for LSPE is the suppression of systematic effects at sub-μK level. The design of the instruments incorporate multiple modulation and differential processing, and the Arctic night flight ensures minimal environmental contamination. Furthermore, the two instruments feature a common channel in W band as a powerful check for systematic effects. In addition to aiming at frontier science, the development of LSPE's large cryogenic polarimeter arrays of bolometers and coherent receivers represents a contribution to the technological development in the field of CMB polarization instruments.

## 7. ACKNOWLEDGMENTS

We gratefully acknowledge support from the Italian Space Agency through contract I-022-11-0 "LSPE".

# REFERENCES

[1] Kovac, J. M., Leitch, E. M., Pryke, C., *et al.,* "Detection of polarization in the cosmic microwave background using DASI", Nature 420, 772-787 (2002)

[2] Readhead, A. C. S., Myers, S. T., Pearson, *et al.,* "Polarization Observations with the Cosmic Background Imager", Science, 306, 836-844, (2004)

[3] Montroy, T. E., Ade, P. A. R., Bock, J. J., *et al.,* "A Measurement of the CMB <EE> spectrum from the 2003 flight of BOOMERANG", The Astrophysical Journal, 647, 813, (2006)

[4] Spergel, D. N., Verde, L., Peiris, H. V., *et al.,* "`First year Wilkinson Microwave Anisotropy Probe (WMAP) observations: Determination of cosmological parameters", The Astrophysical Journal Supplements, 148, 175, (2003)

[5] QUaD collaboration: Pryke, C., Ade, P., *et al.,* "Second and third season QUaD CMB temperature and polarization power spectra," Astrophys. J. 692 (2009)1247

[6] Chiang, H. C., Ade, P. A. R., Barkats, *et al.,* "Measurement of CMB Polarization Power Spectra from Two Years of BICEP Data," The Astrophysical Journal, 711, 1123, (2010)

[7] Planck Collaboration, Ade, P. A. R., Aghanim, N., Arnaud, M., *et al.,* "Planck early results. I. The Planck mission" Astronomy & Astrophysics, 536, A1, 2011

[8] Fraisse, A. A., Ade, P. A. R., Amiri, *et al.,* "SPIDER: Probing the Early Universe with a Suborbital Polarimeter", arXiv:1106.3087 [astro-ph.CO] (2011).

[9] Reichborn-Kjennerud, B., Aboobaker, A. M., Ade, P., *et al.,* "EBEX: a balloon-borne CMB polarization experiment". Proc. SPIE 7741, 77411C (2010)

[10] The Polarbear Collaboration: Errard, J., Ade, P. A. R., Anthony, A., *et al.,* "The new generation CMB B-mode polarization experiment: POLARBEAR", arXiv:1011.0763 [astro-ph.IM].

[11] Seljak, U., and Zaldarriaga, M.,"Signature of gravity waves in polarization of the microwave background", Phys. Rev. Lett. 78, (1997), 2054

[12] Lyth, D. H., "What would we learn by detecting a gravitational wave signal in the cosmic microwave background anisotropy?", Phys. Rev. Lett. 78, (1997), 1861

[13] Liddle, A. R., "The Inflationary energy scale", Phys. Rev. D 49 (1994) 739

[14] Tegmark, M., "What does inflation really predict?", JCAP 0504 (2005) 001

[15] Boyle, L. A., Steinhardt, P. J., and Turok, N., ``Inflationary predictions reconsidered'', Phys. Rev. Lett. 96, (2006), 111301

[16] de Putter, R., Zahn, O., and Linder, E.V., "CMB Lensing Constraints on Neutrinos and Dark Energy", Phys. Rev. D 79 (2009) 065033

[17] Lesgourgues, J., Perotto, L., Pastor, S., and Piat, M., "Probing neutrino masses with CMB lensing extraction", Phys. Rev. D 73 (2006) 045021

[18] Smith, K. M., Cooray, A., Das, S., *et al.,* "CMBPol Mission Concept Study: Gravitational Lensing", AIP Conf. Proc. 1141 (2009) 121

[19] Bersanelli, M., Mennella, A., Morgante, *et al.,* "A coherent polarimeter array for the Large Scale Polarisation Explorer balloon experiment", these proceedings (2012)

[20] de Bernardis, P., Aiola, S., Amico, G., *et al.,* "SWIPE: a bolometric polarimeter for the Large-Scale Polarization Explorer", these proceedings (2012).

[21] Page, L., Hinshaw, G., Komatsu, E., *et al.,* "Three-Year Wilkinson Microwave Anisotropy Probe (WMAP) Observations: Polarization Analysis", The Astrophysical Journal Supplement, 170, 335-376 (2007).

[22] Gold, B., Odegard, N., Weiland, J. L., *et al.,* "Seven-year Wilkinson Microwave Anisotropy Probe (WMAP) Observations: Galactic Foreground Emission", The Astrophysical Journal Supplement, 192, 15 (2011).

[23] Ponthieu, N., Macías-Pérez, J. F., Tristram, M., *et al.*, "Temperature and polarization angular power spectra of Galactic dust radiation at 353 GHz as measured by Archeops", Astronomy and Astrophysics, 444, 327-336, (2005)

[24] Macellari, N., Pierpaoli, E., Dickinson, C., Vaillancourt, J. E., "Galactic foreground contributions to the 5-year Wilkinson Microwave Anisotropy Probe maps", Monthly Notices of the Royal Astronomical Society, 418, 888-905, (2011)

[25] Leach, S. M., Cardoso, J.-F., Baccigalupi, *et al.,* "Component separation methods for the PLANCK mission", Astron. & Astrophys., 491, 597-615 (2008)

[26] Bonaldi, A., Ricciardi, S., "Forecast of B-mode detection at large scales in the presence of noise and foregrounds", Monthly Notices of the Royal Astronomical Society, 414, 615–620, (2011)

[27] Smith, I. S. Jr., "The NASA Balloon Program: looking to the future", Advances in Space Research, 33, 1588–1593 (2004)

[28] Chang, J., Adams, J. H., Ahn, H. S., *et al.,* "An excess of cosmic ray electrons at energies of 300–800 GeV", Nature, 456, 362-365 (2008)

[29] Pascale, E., Ade, P. A. R., Bock, J. J., *et al.,* "The Balloon-Borne Large Aperture Submillimeter Telescope: BLAST" The Astrophysical Journal, 681, 400-414 (2008)

[30] de Bernardis, P., Ade, P. A. R., Bock, *et al.,* "A flat Universe from high-resolution maps of the cosmic microwave background radiation", Nature, 404, 955-959 (2000)

[31] Peterzen, S., Masi, S., Dragoy, *et al.,* "Long Duration Balloon flights development (Italian Space Agency)", Mem. S. A. It., 79, 792-798 (2008)

[32] Peterzen, S., Masi, S., private communication, 2012.

[33] Macías-Pérez, J. F., Lagache, G., Maffei, B., *et al.,* "Archeops In-flight Performance, Data Processing and Map Making", Astronomy and Astrophysics, 467, 1313-1344 (2007)

[34] Mennella, A., Bersanelli, M., Butler, R.C. et al, "Planck early results. III. First assessment of the Low Frequency Instrument in-flight performance", A&A, 536, A3 (2011)

[35] Mennella, A., Bersanelli, M., Butler, R.C. et al, "Planck pre-launch status: Low Frequency Instrument calibration and expected scientific performance", A&A, 520, A5 (2010)

[36] Mennella, A., Villa, F., Terenzi, L, et al, "The linearity response of the Planck-LFI flight model receivers", JINST, 4, T12011 (2009)

[37] Planck HFI Core Team, Ade, P. A. R., Aghanim, N., Ansari, R., et al., "Planck early results. IV. First assessment of the High Frequency Instrument in-flight performance", Astronomy & Astrophysics, 536, A4 (2011)

[38] Fiorineschi L, Boscaleri A, Baldi M, Rissone P. "A new approach in designing of stratospheric platforms aimed at multi-user LDB flights", in 20th ESA Symposium on European Rocket and Balloon Programmes and Related Research, 22-26 May 2011, Hyères, France, ESA SP-700, 593-598 (2011).

[39] Boscaleri, A., Venturi, V., Tronconi, A., Colzi, R., "A time domain design technique for high precision full digital pointing system in balloon-borne remote infrared sensing" in Acquisition, tracking, and pointing IV; SPIE Meeting, Orlando, FL, Apr. 19-20, 1990 Proceedings of the SPIE, 1304, 127-137, (1990)

[40] Boscaleri A., Venturi, V., Tirelli D., "The ARGO experiment pointing system as an example for other single axis platform pointing systems", Measurement Science and Technology, 5, 190-196 (1994)

[41] Boscaleri, A., Pascale, E., "Attitude Control System for Balloon-Borne Experiments", AIP Conference Proceedings 616, 56-58 (2002).

[42] Rotini, F., Boscaleri, A., Baldi, M., *et al.,* "Planning and optimisation of the stratospheric gondola project: search for a standard", Proceedings of 17[th] Symposium on European Rocket and Balloon Programmes and Relates Research, Sandefjord, Norway, 30 May- 2 June 2005, ESA SP-590, 419-424 (2005)

[43] Boscaleri A., Castagnoli F., Rissone P., et al., "Stratobus: A Multiuser Platform System for Making access to LDB Flight Easier and Cheaper", Proceedings of 19[th] Symposium on European Rocket and Balloon Programmes and Relates Research, ESA-SP-671, 209-213 (2009)

[44] Nati, F., de Bernardis, P., Iacoangeli, A., *et al.,* "A fast star sensor for balloon payloads", Review of Scientific Instruments, 74, 4169-4175 (2003)

[45] Benoît, A., Ade, P., Amblard, A., *et al.,* "Archeops: a high resolution, large sky coverage balloon experiment for mapping cosmic microwave background anisotropies", Astroparticle Physics, 17, 101-124 (2002)

[46] Macías-Pérez, J. F.; Mayet, F.; Aumont, J.; Désert, F.-X., "Global Spectral Energy Distribution of the Crab Nebula in the Prospect of the Planck Satellite Polarization Calibration", The Astrophysical Journal, 711, 417-423 (2010)